\documentclass{article}
\usepackage{graphicx} 
\usepackage{amsmath}
\usepackage{amsfonts}
\usepackage{natbib}
\usepackage{amsthm}

\usepackage[utf8]{inputenc}
\usepackage[T1]{fontenc}
\usepackage{lmodern}
\usepackage{geometry}
\usepackage{dsfont}
\usepackage{setspace}
\usepackage{graphicx}
\usepackage{amsmath, amssymb, amsthm}
\usepackage{booktabs}
\usepackage{caption}
\usepackage{authblk} 
\usepackage{natbib}
\usepackage{hyperref}
\usepackage{multirow}
\usepackage{subcaption}
\usepackage{algorithm}
\usepackage{algpseudocode}
\usepackage{xcolor}
\usepackage{lineno}
\usepackage{tcolorbox}
\usepackage{hyperref}[
  colorlinks=true,
  linkcolor=blue,
  citecolor=blue,
  urlcolor=blue
]
\usepackage{doi}
\usepackage[printonlyused]{acronym}
\usepackage{tikz}
\usetikzlibrary{positioning, arrows.meta, decorations.pathreplacing, shapes.geometric, arrows.meta, positioning, calc, shadows.blur, backgrounds, fit}

\title{Consensus-level substitution rates are distinct from the virion-level rate}
\author[1]{David J. Pascall}
\affil[1]{MRC Biostatistics Unit, University of Cambridge, Cambridge, UK}
\date{}

\begin{document}

\maketitle
\section*{Abstract}
Estimating viral substitution rates is central to evolutionary epidemiology, and recent interest in within-host evolution has sharpened the question of what such rates measure. I distinguish two classes of evolutionary rate estimand that are rarely separated in phylogenetic analysis: the virion-level substitution rate (VLSR), a molecular quantity counting mutational events along lineages, and consensus-level substitution rates (CLSRs), population-summary quantities counting allele frequency threshold crossings from below. CLSRs are indexed by the consensus-generation rule. The VLSR and CLSRs are both biologically meaningful, but not interchangeable. Because the consensus-generation rule defines a given CLSR, it should be reported routinely. This reflection should help analysts make more informed methodological choices when working with sets of virus sequences.

\section*{Introduction}
A major practical task in evolutionary epidemiology is to estimate substitution rates and relate them to epidemiological phenomena, whether at the between-host or within-host level. Interest in within-host evolution has recently increased, with theories that within-host evolution in immunocompromised patients may drive the generation of so-called variants in SARS-CoV-2, echoing an old conjecture that host adaptation could occur stepwise via immunocompromised individuals \citep{doi:10.1086/284937}. Substitution rates are generally estimated by assuming the existence of some kind of molecular clock. However, a molecular clock fitted to population consensus sequences requires careful interpretation, because a consensus can change through shifts in existing allele frequencies without any new mutation. The practical question, therefore, is what biological rate such an analysis naturally estimates and under what conditions it approximates the targetted rate of change along individual viral lineages.  One question that has been alive in the literature recently is how best to measure these rates. Some of the state-of-the-art methods have worked with allele frequencies \citep{10.7554/eLife.56915,Ghafari2025}, while other approaches have used phylogenetic trees and mutations along lineages \citep{Alizon2013,VELASQUEZREYES2025101122,10.1093/ve/veaf065}. The methods cited above do not all estimate the same thing. The studies, explicitly or implicitly, are targetting different biological phenomena. Approaches that fit a model to allele-frequency trajectories target the dynamics of the frequencies themselves. Approaches that count mutations along branches of an inferred phylogeny target the rate of molecular change along lineages, provided the tips are genomes of individuals. A third group, which includes most applied work, fits a phylogenetic model to consensus sequences, and it is this group whose natural estimand is least clear. This reflection focuses on the third case.

\section*{Estimand definitions}

One early definitional separation is between \textit{virion-level data} and \textit{consensus-level data}. In virion-level data, every sequence in the dataset is the genome sequence of a real individual of the viral population. At a given time, a population can have multiple tips because it contains multiple virions.  In consensus-level data, every sequence in the dataset is a consensus sequence generated from a viral population. A consensus sequence is the output of a non-reversible function applied to the population's haplotype distribution; typically, it selects the majority nucleotide sitewise, often with a numerical cut-off below which an ambiguity code is output. As the consensus sequence is a function of the entire viral population, conditional on a fixed consensus generation method, there is a single unique consensus sequence per population at a given time point. This split has immediate consequences. Phylogenetic analysis with virion-level data is essentially a paradigmatic classical phylogenetic analysis with each tip being related by an underlying binary tree (assuming no recombination) and standard phylogenetic intuitions holding. Under this setup, the estimand, which I call the \textit{virion-level substitution rate} (VLSR), corresponds to the normally understood estimand in phylogenetics. The difficulty lies in identifying the natural estimand for phylogenetic analyses with consensus-level data, despite this being standard practice. The relationship between consensus sequences differs from that between sequences from individual virions in two ways.

Firstly, the relationship between consensus sequences is not, in general, binary tree-like, even in the absence of recombination, as has been noted in recent work \citep{10.1371/journal.ppat.1013109}, despite our standard phylogenetic models assuming a binary tree. Consider the standard consensus-generation function, returning the majority nucleotide sitewise. This function maps a distribution over sequences to a single sequence. Every consensus sequence constitutes a summary of the distribution of genetic types in the entire population, so that, assuming the population is well-mixed, if some time $t_1$ is earlier than $t_2$, the consensus sequence at $t_1$ is a direct “ancestor” of that at $t_2$, and, further, is the only “ancestor” of the consensus sequence at $t_2$ by the uniqueness of the output per population. There is no binary tree here. There is just a progression of summary statistics of some time-varying distribution of allele frequencies. However, note that while the relational structure of these consensus sequences is not binary-tree-like, assuming no between-population migration, it is tree-like. To see why, note that the sequence-valued function of the allele frequency distribution is applied to each spatially separated population, where the distributions themselves merge as the process proceeds backwards in time towards the ancestral population that all the spatially distinct populations descend from. The consensus sequence of these distributions therefore follows a non-classical tree-valued process, because, at the time of population merging, two population summaries are replaced by a single summary of their shared ancestral population. Within each population, there is no binary tree, but the populations themselves are linked through one. 

Secondly, the consensus sequence can theoretically exhibit multiple simultaneous changes, both at haplotype replacement and population formation, absent any mutational events. I will consider haplotype replacement first, because it isolates the causal phenomena better. Let there be two viral types in the population, $A$ and $B$, with $n$ SNPs between them, and $B$ growing in frequency. The moment that the majority type changes from $A$ to $B$, the consensus sequence instantaneously changes at the $n$ sites at which they differ. This is a linkage phenomenon. Although the consensus is evaluated sitewise, the allele frequencies at each site are perfectly correlated through their shared genomic architecture. The same process occurs at population formation, that is, both infection and spatial segregation within the host. I will illustrate this with infection from a person with a single viral population, but the argument applies analogously to spatially separated viral populations within a single host. If person $i$ infects person $j$, prior to the time of infection, there is a single consensus sequence in person $i$. At the moment of infection, a second is generated in person $j$ from the consensus rule applied to the infecting viral population. Up to a small perturbation caused by the loss of the virions founding the population in person $j$, the consensus sequence in person $i$ remains constant on transmission. However, the bottlenecking process at infection may lead to a very different set of haplotype, and thus allele, frequencies in person $j$, and thus a large difference between the consensus sequences generated from the viruses in person $i$ and person $j$ immediately after infection. This also implies that the coalescence time of consensus-summary processes must coincide with the spatial split of the populations, because the consensus sequence depends on the distribution and, as such, the second summary doesn't exist until population formation.

The differences in the structure of virion-level data and consensus-level data are explained by these two data types illustrating two different biological processes. Consensus sequences are instantiations of population-level phenomena, while virion sequences are instantiations of individual-level phenomena. These differences in behaviour imply that VLSR and \textit{consensus-level substitution rate} (CLSR) describe different biology and therefore are different estimands. The VLSR is the classical phylogenetic substitution rate, fundamentally a molecular evolutionary quantity. Under full observation, this would be the expected number of mutational events per site per unit lineage-time. To see this, note that if we sample every virion at inactivation and know their ancestry, assuming no multiple mutations at the same site within a generation, we observe every mutation. Hence, the rate of events equals the pure mutation rate. CLSRs are a class of population genetic quantities, depending on the population allele frequencies, indexed by a consensus generation rule, $\phi$, with a member of the class with a particular rule being CLSR$_\phi$. CLSR$_\phi$ is the expected number of times an allele crosses the consensus-calling threshold from below per site per unit population-time. These would not be expected to be the same, and are not even obviously directly comparable.

It is worth stating my formalisation of these two quantities more precisely. Consider a single site, $s$, in a viral population, and let $\mathbf{p}_s(t)$ be the vector of allele frequencies at that site at time $t$, with $p_{s,a}(t)$ the frequency of allele $a$ at site $s$ at time $t$. Because the population is finite, each $p_{s,a}(t)$ takes values in $\{0, 1/N(t), \ldots, 1\}$ and changes by discrete steps. An idealised consensus-generation rule $\phi$ maps the vector of sitewise frequencies to a called state, $c^\phi_s(t) = \phi\big(\mathbf{p}_s(t)\big)$, taking values in the four nucleotides augmented by whatever ambiguity or masking characters the rule returns. In practice, the frequency vector must be estimated from sequence data and the calling rule implemented through a bioinformatic pipeline.
 
Over a specified interval $(0,T]$, let the total number of mutational events at site $s$ in the complete virion genealogy, including events on lineages subsequently lost, be $M_s(T)$.  For the consensus process, the quantity of interest is not the number of times $c^\phi_s$ changes value but the number of times an allele enters the consensus by crossing the threshold from below. We define the number of these events as
\begin{equation*}
\text{ for }a\in\{A,G,C,T\},\text{ }U^\phi_{s,a}(T) = \#\big\{t \in (0,T] : \mathds{1}\{c^\phi_s(t) = a\}
\text{ jumps from } 0 \text{ to } 1\big\},
\qquad
U^\phi_s(T) = \sum_{a} U^\phi_{s,a}(T),
\end{equation*}
with the sum running over the four nucleotides. For a population tree, these entry counts are summed over population branches, with entries at population founding determined by comparison with the parent population’s consensus. Let $L_V(T)$ and $L_P(T)$ denote the total virion-lineage-time and population-time in the respective complete histories, both being measured in units of real time. That is, the tree length of the time trees containing all the virions or populations. The two estimands are
\[
\mathrm{VLSR}
=\frac{\mathbb{E}[M_s(T)]}{\mathbb{E}[L_V(T)]},
\qquad
\mathrm{CLSR}_{\phi}
=\frac{\mathbb{E}[U^\phi_s(T)]}{\mathbb{E}[L_P(T)]},
\]
where expectations are taken over histories generated by the specified biological process. The point of the reflection is that $M_s$ and $U^\phi_s$ count events of different kinds, generated by different parts of the biology, and that neither is a function of the other.
 
The distinction between entries from below and changes for CLSRs is important. At a biallelic site, entries from below and state changes coincide whenever every call is a definite nucleotide. When the rule returns an ambiguity character, including at an exact frequency tie, they can differ. If the rule calls an allele only above a threshold $\tau > 0.5$ and returns an ambiguity character below it, an allele rising above $\tau$ while touching 0.5 on its path produces one entry from below but two changes in the called state (resident to ambiguity, then ambiguity to replacement). Counting entries from below therefore explicitly excludes departures into ambiguity from the event count, while retaining dependence on the consensus rule and the underlying allele-frequency path.

To observe that the VLSR and CLSRs really are different objects, consider, as an extreme illustrative case, a perfectly observed two-type population with standing genetic variation undergoing fluctuating selection with no mutation. At a given allele that differs between the two types, the allele frequency path will be oscillatory. Every time this allele crosses the consensus calling threshold, that allele enters the consensus sequence. Here, because there has been no mutation, and thus no per-lineage state changes, the number of virion-level changes is zero. However, the number of relevant consensus-level changes is the number of times the threshold was crossed from below in the allele frequency path, a number greater than zero. The rates, therefore, must differ, and the CLSR$_\phi$ and VLSR are different objects. By allowing a mutation that emerges and is lost without ever crossing the consensus threshold, we can generate a scenario where the VLSR is greater than the CLSR$_\phi$, so neither rate dominates the other. Of course, in practice, we don't observe the full population, but these definitional differences still stand. Figure 1 shows these scenarios concretely at a single site, tracking both the number of mutations and consensus entries. In the monotone growth example, even though the total count of events is the same, the timing is offset, further illustrating that different phenomena are being tracked.

\begin{figure}[htbp]
\centering
\includegraphics[width=\textwidth]{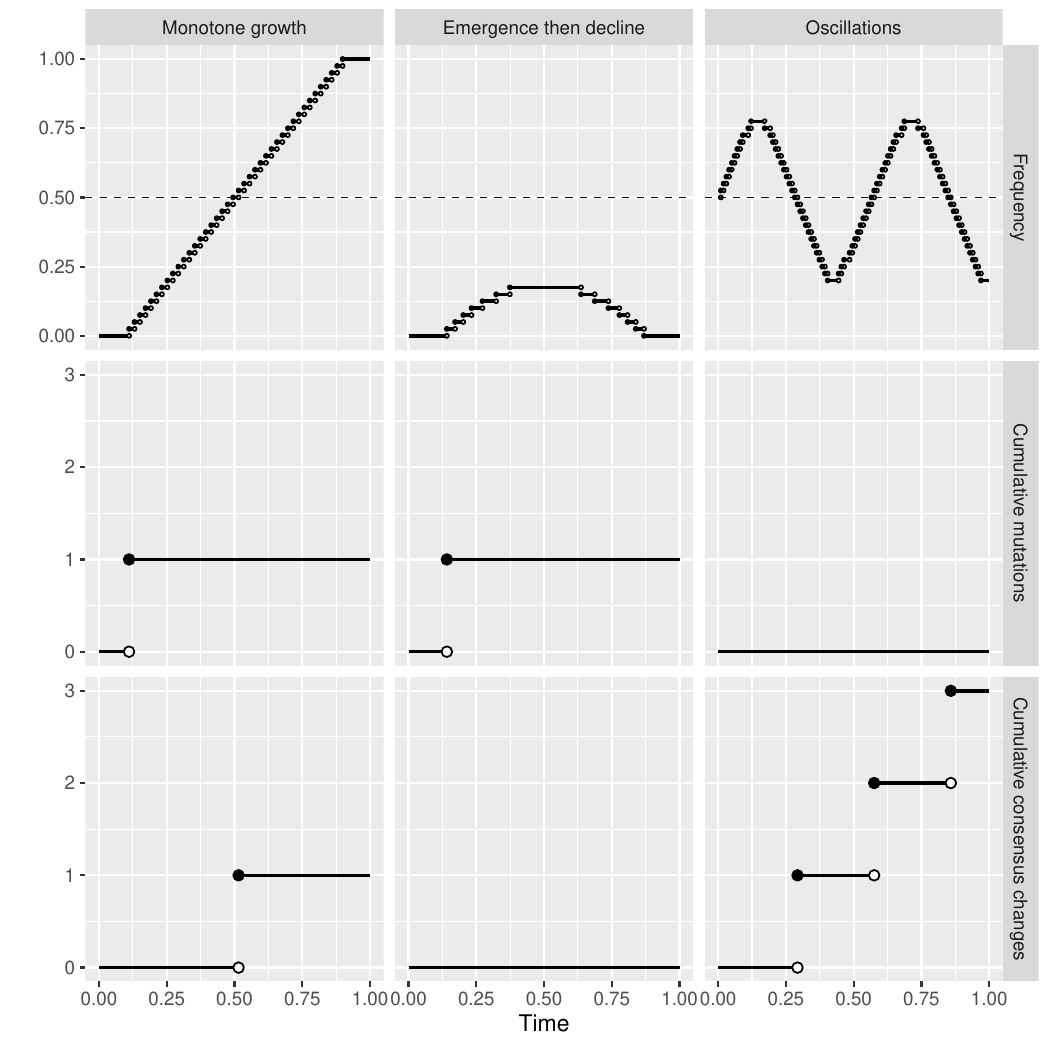}
\caption{Mutational events and consensus changes at a single site under three allele-frequency trajectories with two alleles. $\phi$ is the rule: call the majority allele, return an ambiguity code at ties. Top row: the frequency of a focal allele through time, with the majority-calling threshold for the two allele case shown as a dashed line at 0.5. Middle row: the cumulative number mutations. Bottom row: the cumulative number of crossings of the consensus threshold from below. Filled and open circles denote the value immediately after and immediately before each step. (a) Monotone growth: a single mutation arises and rises to fixation. Both counts reach one, but the consensus change occurs when the allele crosses the threshold, substantially later than the mutational event. (b) Emergence then decline: a mutation arises, reaches a frequency of 0.175, and is lost. One mutational event occurs and no consensus change occurs, so the VLSR exceeds $\mathrm{CLSR}_\phi$. (c) Oscillations: two alleles segregate as standing variation under fluctuating selection with no new mutation. No mutational event occurs, and the consensus changes three times, so $\mathrm{CLSR}_\phi$ exceeds the VLSR. Due to the two allele structure and symmetry, every crossing of the threshold from above is also a crossing of the threshold from below in this case. For the definite nucleotide calls shown here, each entry coincides with a state change; this equivalence need not hold when ambiguity calls intervene.\\
\textbf{Alt text: }Allele frequency trajectories under three scenarios. Monotone growth, emergence then decline and oscillations. The top row of panels shows the allele frequencies. The second row of panels shows the number of mutations. The third row of panels shows the number of crossings of the consensus threshold from below.}
\label{fig:scenarios}
\end{figure}

\section*{Estimand estimation}

We therefore have two estimands, the VLSR and CLSR$_\phi$, which practitioners estimate, each formalising different biological questions. In this section, I focus on the practical questions of estimation of these targets.

Firstly, we need to clarify whether these objects are, in practice, estimable. Both the VLSR and CLSRs are idealised estimands, and the full-observation clause in their definitions is doing heavy lifting. For example, while the VLSR equals the mutation rate if every virion is observed, the estimand is not identifiable from real data without additional assumptions about selection, because non-viable and viable but low-fitness viral particles are lost from the population preferentially between sampling events, so the observed number of mutations underestimates the true number. To put it another way, the VLSR estimate as calculated through real data generated by a plausible observation scheme can be “contaminated” by selective forces, and the sampling scheme itself partially determines the nature of that contamination. This contamination is scientifically interesting in its own right and ties into observations about time-dependent rates, since evolution can drive different kinds of genomic changes over different timescales. However, I resist the urge to generalise the VLSR and family of CLSRs to families including time in the same way that I define a family of CLSRs because of the complexity involved caused by the interaction of the total evolutionary time experienced by the population under study and the sampling regime. I am not claiming that such a generalisation does not fit with how scientists conceptualise these problems, just that it is beyond the scope of this reflection. On the other hand, defining $\phi$ for the CLSRs is necessary, as even conditional on a process and timescale. The estimand is incompletely defined without that anchoring, with special cases for $\phi$ being particularly common choices or of particular biological interest (for instance, CLSR$_\phi$ with $\phi$ calling a nucleotide only when its frequency is 1 is a way to formalise the rate of fixations). 

It is then useful to categorise data and models for estimating these estimands. As would be expected, virion sequences are appropriate for the estimation of the VLSR and consensus sequences are appropriate for the estimation of the CLSR$_\phi$. An important special case is when every consensus sequence generated is also the sequence of a virion in the population; that is, the consensus sequences are not chimeric. In this case, the set of consensus sequences is appropriate for estimation of both the VLSR and CLSR$_\phi$.

These data types can arise through different sequencing workflows. Short-read sequencing of a mixed viral population typically generates reads from fragments or short amplicons originating from different genomes. Read counts at each site estimate allele frequencies, from which a population consensus can be generated using a consensus-calling rule. Long-read sequencing can preserve linkage across a larger portion of an individual molecule, potentially spanning a complete genome or genome segment. However, pooling long reads from different genomes to call a consensus still produces a population summary.

Another consideration is whether amplification starts with material from a mixed population or an individual template. Individual virions can be physically isolated before lysis and amplification \citep{10.1371/journal.pone.0017722}, and extracted nucleic acids can instead be diluted or partitioned for amplification from individual templates, following the principle used in digital PCR or SIP-seq \citep{Sun2022,doi:10.1128/jvi.01137-15}. In the latter cases, the isolated unit is a nucleic-acid template, not necessarily an intact virion. Where amplification begins with a single template or virion and its products remain separately identifiable, subsequent fragmentation and short-read sequencing can still yield an estimate of that initial genome. For viruses amenable to plaque purification, amplification from an isolated plaque provides another route to material representing an individual viral lineage \citep{DePew2013}. Although the sequenced material is then a population produced by replication in culture, its consensus can approximate the founding virion's genome, provided the plaque had a single founder and subsequent mutation and selection have not appreciably altered the consensus. Although these estimates may themselves be obtained by taking a consensus across reads, those reads represent amplified copies of a single starting template. This differs from a population consensus, which summarises variation across the sampled viral population. Thus, neither read length nor consensus calling alone determines whether a sequence represents an individual genome or a population summary. The key questions are therefore whether each reported sequence plausibly represents the genome of an individual virion, and whether sitewise variation among reads reflects diversity in the population being studied. The different data and model types suggested for estimating each estimand are presented in Table 1.

Given appropriate data, a model should be used that coherently estimates the particular estimand. Standard phylogenetic approaches can estimate the VLSR from virion-level data on a tree. These standard tools do not, in general, describe the process generating a CLSR exactly, although they may provide useful approximations. Standard phylogenetic tools assume evolution on a tree following a separate continuous-time Markov chain (CTMC) per site. Such a model describes nucleotide-state changes occurring at random times along branches, with transition rates determined by the current state, and each site acting independently. These assumptions are problematic for estimation of CLSRs: the relation between consensus samples within a well-mixed subpopulation, as is the case in a single within-host viral population, is not binary-tree-like, and simultaneous state changes at internal nodes, as can occur at population formation, almost surely do not occur under standard sequence-evolution CTMCs. This latter case is doubly problematic: the change is concentrated at a single predefined instant, which has probability zero under a process that accrues change along branches of positive length, and it is simultaneous across multiple sites, which has probability zero under site independence.

\begin{table}[htbp]
\centering
\small
\setlength{\tabcolsep}{4pt}
\renewcommand{\arraystretch}{1.15}
\caption{Data types, the estimands they can reasonably support, and the principal
assumptions involved. Instantiation states that each
consensus sequence is also the sequence of some virion in the
population.\\
\textbf{Alt text: }Different data types and how they support the estimation of different estimands.}
\label{tab:datatypes}
\begin{tabular}{@{}p{2.4cm}p{3.0cm}p{2.6cm}p{2.6cm}p{3.4cm}@{}}
\toprule
\textbf{Data type} & \textbf{What each datum is} & \textbf{Estimand
supported} & \textbf{Suitable model} & \textbf{Principal assumptions and
failure modes} \\
\midrule
 
Population consensus sequences &
Output of $\phi$ applied to the population at one time point &
$\mathrm{CLSR}_\phi$; VLSR under instantiation assumption &
Sampled-ancestor genealogy ideally with a sequence model permitting simultaneous
change at founding for CLSR$_\phi$; Standard phylogenetics for VLSR &
$\phi$ must be reported; instantiation generally unverifiable\\
\addlinespace
 
Sitewise allele frequencies &
Per-site frequency estimates from read counts &
Rates of allele-frequency change; $\mathrm{CLSR}_\phi$ for any $\phi$; VLSR after consensus generation under instantiation assumption&
Models for frequency trajectories for rates of allele frequency change; Sampled-ancestor genealogy ideally with a sequence model permitting simultaneous
change at founding for CLSR$_\phi$; Standard phylogenetics for VLSR&
Error and amplification artefacts set a detection floor;linkage information lost \\
\addlinespace
 
Reconstructed haplotypes &
Model-based estimates of genomes present in the population &
VLSR &
Standard phylogenetics, ideally propagating reconstruction uncertainty &
Rare haplotypes poorly recovered; errors correlated across sites;
uncertainty rarely propagated \\
\addlinespace
 
Single-template or single-genome sequences &
Sequence derived from amplification of one template or one plaque &
VLSR &
Standard phylogenetics &
The sequenced object is a template, not necessarily a virion; plaque
consensus approximates the founder only under a single founder and
limited evolution in culture \\
\addlinespace
 
Between-host consensus sequences &
One consensus per host, each a population summary &
$\mathrm{CLSR}_\phi$ at the host level; VLSR under instantiation assumption &
Standard tree priors ideally with a sequence model permitting simultaneous
change at founding for CLSR$_\phi$; Standard phylogenetics for VLSR &
Discontinuity at transmission\\
\bottomrule
\end{tabular}
\end{table}

\subsection*{What are the practical consequences of this for estimation of the VLSR and CLSR$_\phi$?}

If the aim is to estimate the VLSR, standard phylogenetic machinery is appropriate. The data should be either virion-level sequences or consensus sequences, assuming each consensus sequence is instantiated in a virion somewhere in the population. This interpretation still depends on the selection and sampling assumptions discussed above. Note that the instantiation assumption generally cannot be verified from unphased short-read data alone. Partial linkage information can support the co-occurrence of consensus alleles within particular genomic regions, but it does not establish that a single viral genome contains the complete consensus sequence.

If the aim is to estimate the CLSR$_\phi$, consensus sequences provide observations of the relevant population summary, but their analysis requires a model for how that summary changes.  Within each population, there is a single “chain of descent”. Sampled ancestor models (e.g. SA; \cite{Gavryushkina2014}) allow for generalised trees of this form, where an observation can lie on the branch leading to later observations, rather than each observation having to be a separate tip. This gives an appropriate representation of the branching and sampling structure for CLSR$_\phi$ estimation. Whether such a model is identifiable from consensus data is a separate question from whether it is correctly specified. With one consensus sequence per population per time point and few variable sites, the times of population founding are weakly informed, and the model can fit a chain of descent through samples that do not in fact lie on one.
 
The simultaneous changes in the consensus that occur when a haplotype becomes newly dominant potentially occur at every population-founding event, so a representation containing more founding events contains more instants at which the sequence model must accommodate change to which it assigns probability zero. Sampled ancestor models supply an appropriate genealogy, but they do not make the substitution model appropriate. That must be addressed separately. Representing these discontinuities would require a sequence model that permits correlated changes across sites and allows the founding population’s consensus to differ from its parent’s at the founding event. Allowing simultaneous changes along branches addresses the first requirement, but not the second. While the sequence evolution model is misspecified due to this discontinuity, it seems plausible that this may cause only a small bias in many settings. As the new population is founded from a sample of the old population, if the source population has little genetic variation, it seems likely that most of the time the mode of the founder population will be close to the mode of the source population, so the size of the discontinuity may be small and tolerable by standard phylogenetic CTMCs. Conversely, under majority calling, an allele at 40\% in the donor can be at 100\% in the founding population, changing the consensus without any new mutation. However, ultimately, the bias caused by these discontinuities is an empirical question and may change on a virus-to-virus basis.

These differences may matter more in prolonged or mixed infections with divergent haplotypes than in acute infections with little standing variation. Population structure also matters. For example, in a human influenza A(H1N1) infection, upper- and lower-airway samples were found to have different allele-frequency distributions \citep{doi:10.1128/msphere.01043-20}. Changing the sampled compartment can therefore produce a consensus difference even when neither compartment changes. When migration connects populations, the population-tree representation also oversimplifies the demographic processes, and more complex models may be needed to capture important parts of the underlying biology.

Because $\mathrm{CLSR}_\phi$ is defined only relative to $\phi$, it is necessary to be explicit about both the rule $\phi$ and how it relates to the generation of the sequence data. In its idealised form, $\phi$ defines the consensus from the underlying population allele frequencies. Working with real sequencing data additionally involves estimating those frequencies and applying read-depth requirements, quality filters and masking rules. Sequencing and amplification errors, contamination and finite template sampling can distort the estimated frequencies, creating or obscuring apparent threshold crossings. Therefore, in practice the rule that generates a consensus sequence includes at least: the frequency threshold for calling the majority allele; the threshold below which an ambiguity code is returned, and which code; the minimum read depth required for a call and the character returned where depth is insufficient; any masking of sites, primer regions or positions flagged as problematic; any minor-variant cut-off applied before calling; and the treatment of insertions and deletions. These are typically set by a bioinformatic pipeline and inherited by the analyst without comment. The pipeline and version should therefore be reported alongside the frequency thresholds, depth requirements, masking rules, and other settings actually used.

\section*{Discussion}

Hopefully, this reflection clarifies what current methods estimate and supports more informed methodological choices. Both the VLSR and CLSR$_\phi$ are valid estimands, and target different aspects of the evolutionary process. However, they are easily confused, and if the wrong tools or data are used, a setup can be generated where a good estimate of neither is guaranteed. Importantly, the consensus-generating rule is core to the estimand definition for the CLSR$_\phi$, and this should be a non-optional reporting criterion when estimating CLSRs.

\section*{Data availability statement}
This manuscript has no associated data.

\section*{Acknowledgements}
Thanks to Katrina Lythgoe for helpful comments on a draft of this reflection. I used Claude (Anthropic) and ChatGPT (OpenAI) as discussion partners for criticism of the manuscript's framing and argument, for editorial suggestions, initial draft generation of some sections, copy-editing and language refinement.

\section*{Funding}
This research was supported by the National Institute for Health and Care Research (NIHR) Cambridge Biomedical Research Centre (NIHR203312). The views expressed are those of the authors and not necessarily those of the NIHR or the Department of Health and Social Care. This work was supported in part by a grant of access to OpenAI models through the ChatGPT for Academic Researchers program.

\bibliographystyle{apalike}
\bibliography{references}
\end{document}